\documentclass[prl,twocolumn,draft,preprintnumbers,nofootinbib]{revtex4-2}
\usepackage{amssymb}
\begin{document}

\preprint{UPR-1334-T}

\title{Quantization of Black Hole Entropy for Black Holes in Subtracted Geometry}

\author{Mirjam Cveti\v{c}}
\email{cvetic@physics.upenn.edu}
\affiliation{Department of Physics and Astronomy, University of Pennsylvania, Philadelphia, PA 19104, USA,}
\affiliation{Department of Mathematics, University of Pennsylvania, Philadelphia, PA 19104, USA}
\affiliation{Center for Applied Mathematics and Theoretical Physics, University of Maribor, Maribor, Slovenia.}
\author{Malcolm J. Perry}
\email{mjp1@cam.ac.uk}
\affiliation{Centre for Theoretical Physics, School of Physical and Chemical Sciences, Queen Mary University of London, 
327 Mile End Road, London E1 4NS, UK,} 
\affiliation{Department of Applied  Mathematics and Theoretical Physics, Centre for Mathematical Sciences, Wilberforce Road, Cambridge, CB3 0WA, UK}
\affiliation{Trinity College, Cambridge, CB2 1TQ, UK.}
\date{\today}

\begin{abstract}
We carefully examine the exact analytic spectrum of quasinormal modes of general black holes
 in the so-called subtracted
geometry of maximally supersymmetric supergravity. These black holes have the same area and 
surface gravity at both the
outer and inner horizons as the original asymptotically flat black holes. 
We proceed to explore the relationship 
with conformal field
theories that describe horizon physics of these black holes. 
As a consequence, we show that the horizon area of these
black holes is quantized in units of $8\pi l_{Pl}^2$.
\end{abstract}

\maketitle

 Fifty years ago, Hawking \cite{Hawking:1975vcx} showed that the entropy of a stationary black hole
 is determined solely by the area $A$ of its event horizon. To be precise, 
 \begin{equation}
 S = \frac{kA}{4l_{Pl}^2}
 \label{eq:bhent}
 \end{equation}
 where $l_{Pl}$ is the Planck length defined by $l_{Pl}^2 = G\hbar/c^3$.
 Such a formula had previously been conjectured by Bekenstein \cite{Bekenstein:1972tm,Bekenstein:1973ur} although he was
 not able 
 to find the constant of proportionality in (\ref{eq:bhent}). That this is a non-perturbative
 quantum gravity result can be seen from the appearance of a factor of $l_{Pl}^2$ in the denominator 
 of (\ref{eq:bhent}).
The Boltzmann view of entropy \cite{Boltzmann:1877} is that it is the logarithm of the 
 density of states
 of a system. Taking this to be universally true, (\ref{eq:bhent}) should yield information about
 the fundamental nature of both black holes and quantum gravity. It has been speculated 
 \cite{Bekenstein:1974jk,Mukhanov:1986me,Bekenstein:1995ju,Hod:1998vk} that the 
 spectrum of states of a black hole should result in the area of the event horizon being quantized
 in some multiple of $l_{Pl}^2$. We will find some evidence in support of this conjecture with the quantum of area
 being $8\pi l_{Pl}^2$. The same result for area quantization was found in a particular limit for the
 Schwarzschild solution by Maggiore, \cite{Maggiore:2007nq}.
 From here on, we will use natural units such that $G=c=\hbar=k=1$.

 One way of exploring the spectrum of black hole states is through resonance scattering. In many ways, one
 can view a black hole as a gravitational analog of the atomic nucleus. In nuclear physics, information about 
 nuclear structure can be obtained through resonance scattering. It is therefore plausible to suppose the
 same is true for black holes. Resonances can be found by looking at solutions of the differential equation that governs 
 perturbations of the black hole. Physically, one is interested in waves that are ingoing at infinity and bounded on
 the outer horizon. It is these waves that will put the black hole into some excited state which will then decay. 
 Waves obeying these boundary conditions are termed quasinormal modes. 
 In general, they will form a discrete set with 
 time dependence of the form $e^{-i\omega t}$ with $\omega$ lying in the lower half of the complex plane
 and where $\omega$ determines both the energy of the excited state and the timescale for its decay. 
 Ideally, one would like a complete analysis of the quasinormal modes of the Kerr black hole.
 Kerr quasinormal modes have been studied extensively but are only known numerically at present. 
 However, there is a family of black 
 holes that in many ways resemble the Kerr black hole. 
 These black holes are however not solutions of the vacuum field equations for general relativity but rather
 solutions in $N=2$ supergravity to which has been added
 three vector supermultiplets.  In this theory, there are black holes that, rather than living in an asymptotically flat 
 spacetime, live in an asymptotically conical spacetime. Thus the environment of these
 black holes 
 is different to those described by the Kerr metric. The spacetime of these black holes is conventionly
 referred to as subtracted geometry. 
 These spacetimes 
 were introduced in \cite{cvetic:2011dn} and further analyzed in \cite{cvetic:2012tr,cvetic:2013lfa,cvetic:2014ina,cvetic:2014nta,an:2016fzu}.
 For black holes in subtracted geometry, all the quasinormal modes can be found analytically
 which allows us to explore the excited states these black holes. We then relate the spectrum of quasinormal
 modes for black holes in subtracted geometry to the CFT description of horizon physics. 
 The properties of a black hole 
 should not depend on its environment and 
 so we expect that our results will have general applicability. 
 
 \section{Subtracted Geometry}
 
 The metric of our black holes in Boyer-Lindquist type coordinates is
 \begin{eqnarray}  ds^2 = -\Delta^{-\frac{1}{2}}\, {\cal G} \,(dt+{\cal A})^2\nonumber\ \ \ \ \ \  \\ +
  \Delta^{\frac{1}{2}}\Bigl(\frac{dr^2}{X} +
 d\theta^2 + \frac{X}{G}\sin^2\theta d\phi^2\Bigr), \label{eq:subtracted}
 \end{eqnarray}
 where
 \[ \Delta = 4m^2(2mr(\Pi_c^2-\Pi_s^2) + 4m^2\Pi_s^2 - a^2(\Pi_c-\Pi_s)^2\cos^2\theta),  \]
 \[ {\cal G} = r^2 - 2mr + a^2\cos^2\theta,  \ \ \  X = r^2 -2mr + a^2, \]
 \[ {\cal A} = \frac{2ma\sin^2\theta}{{\cal G}}[(\Pi_c-\Pi_s)r+2m\Pi_s]d\phi. \]
and depends on four parameters, $m,a,\Pi_c$ and $\Pi_s$ 
\footnote{Note that
the corresponding asymptotically flat black holes, first found in \cite{Cvetic:1996kv,Chong:2004na},
are parametrised by
six parameters, $m$, $a$, and $\delta_i$ with $i=\{1,2,3,4\}$ which specify the mass 
$M=\frac{m}{4}\sum_{i=1}^4\cosh 2\delta_i$, angular momentum $J=-ma(\Pi_c-\Pi_s)$ 
and four charges $Q_i = \frac{m}{4}\sinh 2\delta_i$. Here $\Pi_c=\Pi_1^4\cosh\delta_i$ and $\Pi_s=\Pi_1^4\sinh\delta_i$. 
The Kerr black hole corresponds to $\delta_i=0$ and thus $\Pi_c=1$ and $\Pi_s=0$. The Kerr-Newman
black hole to $\delta_i=\delta$ and $\Pi_c=\cosh^4\delta$ and $\Pi_s=\sinh^4\delta$. The geometry of 
asymptotically flat black holes differs from the subtracted geometry (\ref{eq:subtracted}) only in that is has a more
complicated warp-factor $\Delta$. The asymptotically flat black holes therefore have the same 
area and surface gravity on both the inner and outer horizons as those in subtracted geometry.}.

 To be a non-extremal black hole requires
$m>\vert a \vert \ge 0$ and $\Pi_c\ge \Pi_s \ge 0.$ The subtracted Kerr geometry corresponds to taking
$\Pi_c=1$ and $\Pi_s=0$ in the expressions above. The geometry of the Kerr metric differs from the 
subtracted Kerr one only simply by the replacement of
$\Delta$ 
 by $\Delta_{Kerr}$  with
\[ \Delta_{Kerr} = (r^2 + a^2\cos^2\theta)^2. \]
Just like the Kerr metric, these black holes are stationary and axisymmetric, that is to say  
independent of both $t$ and $\phi$. Also, just as in Kerr,  the
zeros of $X$ at
\[ r_\pm = m \pm \sqrt{m^2-a^2} \] 
are the inner and outer horizons with the same value of the area and surface gravity as for the 
Kerr black hole. 

\section{Thermodynamics}

The subtracted black holes are completely characterised by a set of charges determined by the 
parameters of the
solution.  We give explicit formulae for these charges \cite{an:2016fzu}:
the black hole mass $M$ is 
\[M = \frac{1}{4}m(\Pi_c^2+\Pi_s^2) - \frac{a^2}{8m}(\Pi_c-\Pi_s)^2, \]
its angular momentum $J$ is
\[ J = -ma(\Pi_c-\Pi_s),\]
and in addition the gauge fields yield both  an electric charge $Q^{(e)}$ and a magnetic charge $Q^{(m)}$
\[ Q^{(e)} = \frac{2m^2\Pi_c \Pi_s + a^2(\Pi_c-\Pi_s)^2}{4m}, \ \ \ \  Q^{(m)} = \frac{3m}{2}. \]
The entropy $S_\pm$ of the outer and inner horizons is given by the area formula (\ref{eq:bhent}). Thus
\[S_\pm = 2\pi m(\Pi_c r_\pm + \Pi_s r_\mp). \]
Knowledge of the entropy motivates a complete exploration of the black hole thermodynamics.
The temperatures of the outer and inner horizons are determined, as usual, by the surface gravity 
on the horizons and 
are given by
\[
T_\pm = \pm \frac{(\Pi_c^2 - \Pi_s^2)\sqrt{m^2-a^2}}{4\pi m (\Pi_c r_\pm + \Pi_s r_\mp)}.\]
It should be noted that $T_-<0$
which is perhaps a bit surprising. 
\footnote{Again, the entropy and temperature formulae are the same as those for
the corresponding asymptotically flat black holes. The negative temperature for these geometries was first 
derived in \cite{cvetic:2018dqf}.}

However, this comes about because
$\frac{\partial S}{\partial M} < 0$. An explanation of why this is not paradoxical can be found in 
\cite{Frenkel:2015,cvetic:2018dqf}.
On each of the two horizons, there are potentials conjugate to $J, Q^{(e)}$ and $Q^{(m)}$.
The angular velocities $\Omega_\pm$ are 
\[\Omega_\pm = \frac{a}{4m^2}\frac{(\Pi_c-\Pi_s)(\Pi_c r_\mp + \Pi_s r_\pm)}{\Pi_c r_\pm + \Pi_s r_\mp}
\]
and electric and magnetic potentials on the outer and inner horizons are
\[ \Phi_\pm^{(e)} = \frac{ \Pi_c r_\mp + \Pi_s r_\pm}{\Pi_c r_\pm + \Pi_s r_\mp} \]
and
\begin{eqnarray}  \Phi_\pm^{(m)} =\frac{\Pi_c-\Pi_s}{8m^2}\frac{1}{(\Pi_c r_\pm + \Pi_s r_\mp) } \nonumber \\
\times \Biggl(
(\Pi_c^2-\Pi_s^2)(-4m^3+6ma^2) \nonumber \\
 \mp(\Pi_c^2+\Pi_s^2)\sqrt{m^2-a^2}(4m^2+2a^2) \nonumber \\
 \mp\Pi_c\Pi_s\sqrt{m^2-a^2}(8m^2-4a^2)
  \Biggr). \nonumber
  \end{eqnarray}

From these thermodynamic relationships, we can deduce that both the Smarr relations
\begin{equation}
M = 2T_\pm S_\pm + 2\Omega_\pm J + \Phi_\pm^{(e)}Q^{(e)}+\Phi_\pm^{(m)}Q^{(m)} \end{equation}
and the first law of thermodynamics 
\begin{equation} dM = T_\pm dS_\pm + \Omega^\pm dJ + \Phi_\pm^{(e)}dQ^{(e)}+\Phi_\pm^{(m)}dQ^{(m)}.\end{equation}
hold for subtracted  black holes.

It hs been widely conjectured that this data encodes the thermodynamics of a pair of two-dimensional 
conformal field theories, $CFT_L$ and $CFT_R$, \cite{cvetic:1997xv,strominger:1997eq,cvetic:2009jn,castro:2010fd,
cvetic:2011dn,castro:2013lba,castro:2013kea,haco:2018ske}. The relevant thermodynamic quantities are

\[ T_L = \frac{T_+ T_-}{T_+ + T_-} =  \frac{\Pi_c + \Pi_s}{8\pi m}, \]
\[ T_R = \frac{T_- T_+}{T_- - T_+} = \frac{(\Pi_c-\Pi_s)\sqrt{m^2-a^2}}{8\pi m^2}, \]

\[ S_L = \frac{1}{2}(S_++S_-) = 2\pi m^2(\Pi_c+\Pi_s), \]
\[ S_R = \frac{1}{2}(S_+-S_-) = 2\pi m \sqrt{m^2-a^2}(\Pi_c-\Pi_s). \]
The angular momentum and charges are universal but the corresponding potentials for $CFT_L$ are
\[ \Omega_L =  -\frac{a}{8m^2}(\Pi_c-\Pi_s),  \ \ \ \
\Phi_L^{(e)} = -\frac{1}{2}, \]
\[ \Phi_L^{(m)} = -\frac{1}{4}(\Pi_c+\Pi_s)^2 - \frac{a^2}{8m^2}(\Pi_c-\Pi_s)^2 \]
and for $CFT_R$ 
\[ \Omega_R = \frac{a(\Pi_c-\Pi_s)}{8m^2}, \ \ \ \
\Phi_R^{(e)} = \frac{1}{2}, \]
\[ \Phi^{(m)}_R= (\Pi_c-\Pi_s)^2\Bigl(-\frac{1}{4} + \frac{3a^2}{8m^2}\Bigr). \]
Similarly to the horizons, there are Smarr-type formulae and first laws for both $CFT_L$ and $CFT_R$. 
Let the energy of these $CFT$s be $E_L$ and $E_R$.
Then 
\footnote{It turns out that both the Smarr formula and the first law of thermodynamics for the asymptotically
flat black holes can also be cast in a similar form for both of the $L$ and $R$ sectors where now $M$ is the 
mass of the asymptotically flat black hole and each of the four charges $Q_i$ is accompanied by its corresponding gauge
potential.}
\begin{equation}
E_{L,R}= 2T_{L,R}S_{L,R} + 2\Omega_{L,R} J + \Phi^{(e)}_{L,R}Q^{(e)}+\Phi^{(m)}_{L,R}Q^{(m)},\end{equation}
and 
\begin{equation} dE_{L,R} = T_{L,R} dS_{L,R} + \Omega_{L,R} dJ + \Phi^{(e)}_{L,R}dQ^{(e)}+\Phi^{(m)}_{L,R}dQ^{(m)} \end{equation}
for both $CFT_L$ and $CFT_R$. $E_L=E_R={M/2}$. These formulae relate to the changes in these holographic $CFT$s in response to
dynamical change in the black hole geometry.

\section{Quasi-Normal Modes}

The wave equation for minimally coupled massless scalars that carry no charge
has solutions $\chi$ of the form  
\[ \chi \sim e^{-i\omega t +  ik\phi} P_l{}^k(\cos\theta)R_{lk\omega}(r). \]
For the Kerr metric, both the polar functions and radial functions
are confluent Heun functions which lead to complications in their evaluation. 
\cite{vishveshwara:1970zz,chandrasekhar:1975zza,leaver:1985ax,leaver:1986gd,nollert:1992ifk,nollert:1993zz,
nollert:1999ji,kokkotas:1999bd,motl:2003cd,ferrari:2007dd,berti:2009kk,konoplya:2011qq,CarneirodaCunha:2015hzd,
CarneirodaCunha:2019tia,Aminov:2020yma,Bonelli:2021uvf}. For subtracted geometry, 
the polar functions are just the associated Legnedre polynomials $P_l^k$ and the radial functions
are hypergeometric functions.
The radial equation has two independent solutions. One is 
\[ R_{lk\omega}(r) \sim
(r-r_-)^p(r-r_+)^qF(a,b;c;\frac{r-r_+}{r-r_-}) \]
where 
\[p= i\frac{\omega - k\Omega_+}{4\pi T_+} - (l+1) \ \ \ \
q= -i\frac{\omega - k\Omega_+}{4\pi T_+},\]
\[ a=1+l-i\frac{\omega}{4\pi T_R}+i\frac{k\,\Omega_+}{2\pi T_+}. \ \ \ \ 
b=1+l-i\frac{\omega}{4\pi T_L} \]
and
\[ c = 1-i\frac{\omega}{2\pi T_+} +i\frac{k\Omega_+}{2\pi T_+} \]
and which represents waves that are ingoing on the future outer horizon. 
The second solution represents waves that are outgoing on the past outer horizon and are 
therefore not relevant to the present discussion, \cite{cvetic:2013lfa,cvetic:2014ina}.

To find the form of $R_{lk\omega}(r) $ for large $r$, we can use the connection formula 
\begin{eqnarray}
&F(a,b;c;\frac{r-r_+}{r-r_-}) = \nonumber \\
&\frac{\Gamma(c)\Gamma(c-a-b)}{\Gamma(c-a)\Gamma(c-b)}F(a,b;a+b-c+1;\frac{r_+-r_-}{r-r-})  \nonumber \\
&+\Biggl\{\Bigl(\frac{r_+-r_-}{r-r_-}\Bigr)^{c-a-b}\frac{\Gamma(c)\Gamma(a+b-c)}{\Gamma(a)\Gamma(b)} \nonumber \\
&F(c-a,c-b;c-a-b+1;\frac{r_+-r_-}{r-r-})\Biggr\}.\label{eq:connection}
\end{eqnarray}
As $r\rightarrow\infty$ the first term in 
(\ref{eq:connection}) results in a contribution to $R_{lk\omega}(r)$ that scales like $r^{-l-1}$ whereas the 
second term scales like $r^l$.  Requiring $R_{lk\omega}(r)$ to be bounded as $r\rightarrow\infty$
requires the second term to have vanishing coefficient. It is this condition that determines the complex 
frequencies of the quasi-normal modes. 
Since the $\Gamma$-function has poles at negative integers and zero, we require either $a$ or $b$ 
to be a negative integer or zero. There are then two families of quasi-normal modes remarkably corresponding 
precisely to the two CFTs. We term these two families of quasi-normal modes
the $L$ family and the $R$ family. For the left family, we find 
\footnote{These results agree with the calculation in 
\cite{cvetic:2014ina}. In \cite{cvetic:2013lfa} the results are unfortunately missing a factor of $2$.}
\begin{equation} \frac{\omega_L}{T_L} = -4\pi i(n+l) \label{eq:omegal} \end{equation}
and for the right  
\begin{equation} \frac{\omega_R}{T_R} = -4\pi i(n+l)+\frac{k\Omega_+}{T_+} \label{eq:omegar} \end{equation}
where $n \in \mathbb{Z}^+$. Note that in both cases, the imaginary part of $\omega$ is negative
as one would expect for a mode that is damped. It represents the width or lifetime of the state that is being
excited.

The cross-section for absorption by the black hole contains a factor of
\begin{equation} \vert\Gamma(a)\Gamma(b)\vert^2. \end{equation}
This last term should be compared to a corresponding term in the absorption cross-section for a $CFT_L\otimes CFT_R$ 
where the waves have frequencies 
$\omega_L$ and $\omega_R$ and the CFT's have temperatures of $T_L$ and $T_R$:
\begin{equation}
 \bigl\vert \Gamma(1+\frac{i\omega_L}{4\pi T_L})\Gamma(1+\frac{i\omega_R}{4\pi T_R}) \bigr\vert^2. \end{equation}
We see now that there is a precise matchup between the absorption cross-section of the black hole and the
absorption cross-section of a pair of conformal field theories at temperatures $T_L$ and $T_R$ provided
that no angular momentum is added so that $k=0$. Our relations show that at fixed $J, Q^{(e)}$ and $Q^{(m)}$
the energy levels 
of the 
black hole, or equivalently $CFT_L$ and $CFT_R$,  are evenly spaced. Elsewhere, we will explain how to remove the 
constraint of fixed angular momentum without changing our results. Also, we will
explain how to extend these ideas to five-dimensional black holes. 
Our results reveal a remarkable correspondence in that they give considerable
weight to the idea that these conformal field theories are holographic avatars of the true degrees of 
freedom of a 
black hole. .

\section{Oscillators}

It may seem strange that the $L$ family of quasi-normal modes has a frequency that is purely imaginary. 
One should not find this to be a matter of concern for black hole physics as was first clearly explained by
Maggiore \cite{Maggiore:2007nq,Medved:2008iq}. Consider a classical damped oscillator that we take to be a model of
the states of the black hole. Such a system is governed by a familiar second-order ordinary 
differential equation. Modelling the black hole in a way similar to the liquid drop model 
of the nucleus, 
the amplitude of its oscillation will be $f(t)$ and governed by
\begin{equation}
\ddot f + 2\gamma\, \dot f + \omega_o^2\, f = { forcing\ term}. \end{equation}
Suppose the system is initially stationary with $f(t)=0$ for $t<0$ and the forcing term is just $\delta(t)$.
The solution is then 
\begin{equation} f(t) = \frac{i}{\omega_+-\omega_-}\Bigl[\ e^{-i\omega_+t} - e^{-i\omega_-t} \ \Bigr], \end{equation}
where
\begin{equation}  \omega_\pm = \pm \sqrt{\omega_o^2 - \gamma^2} - i\gamma. \label{eq:mod} \end{equation}
Thus we see the damping factor is just $\gamma$ but the frequency of oscillation is $\sqrt{\omega_o^2-\gamma^2}$.
We therefore see that an apparently purely damped quasi-normal mode does indeed correspond to a natural frequency of
oscillation. We conclude that the spectrum of quasi-normal modes, even if resulting in a purely imaginary $\omega$,
does indeed reflect the modes of excitation of the 
black hole and these in turn reflect the spectrum of excitations of $CFT_L \otimes CFT_R$.

\section{ Perturbations}
Suppose we now add to the black hole a quantum of energy $\delta M$ whilst keeping the angular momentum J,
and the charges $Q^{(e)}$ and $Q^{(m)}$ constant. A transformation of this kind could be the result
of the absorption of a quantum by either the left family or the right family of quasinormal modes. If from the
right family, then we are looking at the case where $k=0$.
The energy added will be either $\vert \omega_L  \vert$ or $ \vert \omega_R \vert $ as can be seen 
from equation (\ref{eq:mod})
In either case, one can use the first law for the left family or the right family as appropriate to see that
the change in $S_L$ or $S_R$ is $2\pi \times ({\rm some\ integer})$. Since $S_+ = S_L + S_R$, we conclude 
the area of the outer horizon is quantized in units of $8\pi l_{Pl}^2$.

At this point we need to resolve a puzzle. In \cite{cvetic:2018dqf} it was shown that in the Smarr law for 
$CFT_L$ and $CFT_R$, $E_L=E_R=M/2$. To deduce horizon quantization, we used the first law for $CFT_L$ and $CFT_R$
and took $dE_L= dM = \vert \omega_L \vert $ or $dE_R= dM = \vert \omega_R \vert $. This is a reasonable thing to do 
since the first law describes 
infinitesimal changes, something that is consistent with adding a single quantum in perturbation theory. 
However, there is now tension with the Smarr formulae. However, the Smarr formulae, constitutive relations, relate
to macroscopic quantities
and one expects them to be true only on average after some degree of equilbrium has been achieved. Of course, 
if the black hole absorbs a random collection of quanta over an extended period, then at each step, we expect the
first law to hold. However, we expect the Smarr formulae to hold only on average over the extended period.
\footnote{ The results are true for the whole family of charged subtracted geometries. Of course they reduce to the 
Kerr one for $\Pi_c=1$ and $\Pi_s=0$ and for Kerr-Newman with $\Pi_c=\cosh^4\delta$ and $\Pi_s=\sinh^4\delta$.}

It also possible that beyond
perturbation theory more complicated things happen, but it is remarkable that such simplicity as we have 
found appears. Elsewhere, we will explain how to remove the 
constraint of fixed angular momentum without changing our results. Also, we will
explain how to extend these ideas to five-dimensional black holes. 
Our results reveal a remarkable correspondence. They give considerable
weight to the idea that these conformal field theories are holographic avatars of the true degrees of 
freedom of a black hole.  We also intend to carefully examine the cases of Kerr and Kerr-Newman to investigate
possible similar correspondences.

\begin{acknowledgments}
The work of MC is supported by DOE (HEP)
Award DE-SC0013528, the Simons Foundation
Collaboration grant \#724069, the Slovenian Research
Agency (ARRS No. P1-0306) and Fay R. and Eugene L. Langberg Endowed Chair funds.
MJP is partially supported by
the Science and Technology Facilities Council (STFC) Consolidated Grants ST/T000686/1
“Amplitudes, Strings and Duality” and ST/X00063X/1 “Amplitudes, Strings and Duality”.
we would like to thank the  Mitchell foundation for support and  its hospitality at Cook's Branch.
We would also like to thank Slava Mukhanov, Andrew Strominger and Bernard Whiting for helpful
conversations on this topic.
\end{acknowledgments}

\bibliography{shortsubqnm.bib}

\end{document}